# Non-line-of-sight photoacoustic imaging

Yuting Shen, *School of Information Science and Technology, ShanghaiTech University,* Shanghai, China shenyt@shanghaitech.edu.cn

Xiaohua Feng*, *Research Center for Humanoid Sensing  Zhejiang Laboratory*  Hangzhou, China fengxiaohua@zhejianglab.com

Fei Gao*, *School of Information Science and Technology, ShanghaiTech University* Shanghai, China, gaofei@shanghaitech.edu.cn

*Abstract*—**Photoacoustic imaging is a promising imaging technique for human brain due to its high sensitivity and functional imaging ability. However, the skull would cause strong attenuation and distortion to the photoacoustic signals, which makes non-invasive transcranial imaging difficult. In this work, the temporal bone is selected as an imaging window to minimize the influence of the skull. Moreover, non-line-of-sight photoacoustic imaging is introduced to enhance the field of view, where the skull is considered as a reflector. Simulation studies are carried out to show that the image quality can be improved with reflected signal considered.**

## I. Introduction

Photoacoustic (PA) imaging is an emerging imaging modality that combines the high specificity of optical imaging with the deep penetration of ultrasonic waves [1]. It has been applied in a wide range of scenarios, including small animal imaging [2], breast imaging [3] and so on. Recently, transcranial PA imaging of human brain has attracted great attention and it has great significance for the diagnosis of brain diseases [4]. Compared with traditional techniques, transcranial PA imaging can realize high-resolution reconstruction of blood vessel distribution without adding contrast agent, and the system cost is lower and harmless.

However, the biggest challenge in applying PA imaging to transcranial imaging lies in the presence of skull [5]. The skull would cause strong attenuation and distortion to the acoustic signal. When it comes to human skull, it could reach 1~2 centimeters in thickness. One solution to solve this problem is to select the thinnest area of human skull for the detection of PA signals. T.raham et al. [6]  studied the imaging window of transcranial PA imaging, and found that the acoustic window of sphenoid sinus and eye provided the best image quality. In this work, we propose to image through the temporal bone with non-line-of-sight PA signal detection. The temporal bone  is only 1 millimeter at the thinnest area, thus minimizing the attenuation and distortion of sound waves caused by skull. However it leads to another problem of limited view. Since the location of temporal bone is in the lateral region of the brain, the field of view (FOV) is quite limited.

Non-line-of-sight imaging is a new optical computational imaging technology developed in recent years [7]-[8], which is to image the area beyond the line of sight of the observer. An intermediate plane is always used to image the hidden object that is blocked by other objects. Lindell et al. [9] developed acoustic non-line-of-sight imaging, which uses sound waves to image the hidden objects. The experimental results proved its feasibility. Huang et al. [10] used a 45-deg acoustic reflector in photoacoustic imaging to double the detection coverage. Zhang et al. [11] used two ultrasound reflector to help increase the view angle of handheld linear-array ultrasound probe. However, these works are all based on manually added external reflector to expand the FOV.

In this paper, we propose non-line-of-sight PA imaging technique to improve the detection view by using an intrinsic reflector, the human skull.

## II. METHODS

In this section, we would introduce the implemented reconstruction algorithm of non-line-of-sight PA imaging.

### A. Image reconstruction with reflective PA signals

We use subscript character D to denote detector, use S to denote a point which is considered as source, and use R to denote reflector. We assume there are $N$ real detectors and $M$ reflection points on the reflecting interface. Their positions and normal vectors are given by $P_{D_n}, N_{D_n}, P_{R_m}, N_{R_m}$ respectively.

The received PA signal by a real detector can be expressed as

$$S_{D_n}(t) = \sum_x \sum_y \sum_{R_m} F_R(v_{RS}, v_{DR}, N_{R_m}) F_D(v_{DR}, N_{D_n})$$

$$P_0(x, y, t - (\frac{c_{DR}}{c_0} + \frac{c_{RS}}{c_0})) \quad (1)$$

$F_R(\cdot)$ denotes the reflection function, where $v_{RS}, v_{DR}$ denote the vector from reflector to a source point and the vector from a detector to a reflector, respectively, and $N_{R_m}$ denotes the normal vector of the reflection point $R_m$. $F_D(\cdot)$ denotes the directivity function of the detector.

$$F_R(v_{in}, v_{out}, N) = ((v_{in} - 2(N \cdot v_{in})N) \cdot v_{out})^{k_R} \quad (2)$$

$$F_D(v_{in}, N) = (v_{in} \cdot N)^{k_D} \quad (3)$$

Here $v_{in}, v_{out}, N$ are unit vectors, $k_R, k_D$ are the surface expansion index of the reflecting interface and the probe directivity index.

$P_0$ denotes the initial pressure distribution given a point at position $(x, y)$, which is the object of reconstruction. $c_{DR}$ and $c_{RS}$ denote the distance from real detector to reflector and from reflector to a source point, respectively. $c_0$ is the speed of sound in soft tissue.

The reflection point is considered as a virtual detector, then the received signal $S_{D_n, R_m}$ is

$$S_{D_n, R_m}(t) = \frac{1}{\sum_{R_m} F_D(v_{DR}, N_{D_n})} S_{D_n}(t) \quad (4)$$

Thus, it can be represented as

$$S_{D_n, R_m}(t) = \sum_x \sum_y F_R(v_{RS}, v_{DR}, N_{R_m}) P_0(x, y, t' - \frac{c_{RS}}{c_0}) \quad (5)$$

where $t' = t - \frac{c_{DR}}{c_0}$. Then the non-line-of-sight PA image reconstruction problem is similar with the classic one, which is then implemented by delay and sum (DAS) algorithm.

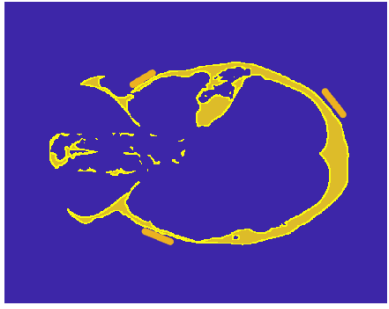

Fig. 1. The skull phantom setting and the position of 3 acoustic array sensors are shown. The top left one is for the reflective reconstruction and the other two are for direct reconstruction.

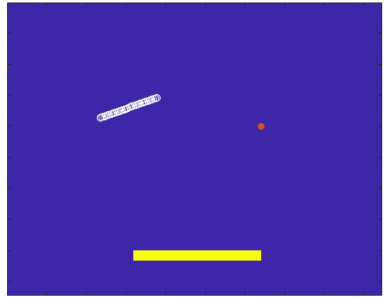

Fig. 2. The phantom setting of simple case validation. A point source and a smooth reflector are set. A linear array is placed towards the direction of reflection.

*B. Image reconstruction superimposed with reflected PA signals*

The reflective signal suffers attenuation when passing through the interface. In addition, the image reconstruction by only using the reflective PA signals also has a limited view problem. To maximize the FOV, we further propose to combine the PA signals detected with and without reflection.

In this way, the overall process includes following steps:

1) The skull structure information is firstly obtained by ultrasound or other imaging modalities.

2) The proper skull section is chosen and divided into the reflection interface. The normal vector of the interface is calculated by local gradient and interpolation.

3) The photoacoustic signal is detected and reconstructed twice. In first reconstruction, it is seen as direct signal and the image is reconstructed with DAS. In second reconstruction, the reflection interface is used and the PA signal is considered as reflective signal.

4) The two images are properly superimposed.

## III. EXPERIMENT SETTINGS AND RESULTS

### A. Human brain phantom generation

In order to create a realistic model of the human skull, we use three-dimensional CT images to create the digital phantom [12]. The skull and brain tissue are labeled by threshold segmentation. In order to further simulate the structure of the skull, we model the skull in three layers [13], including the high-density part of the innermost layer, the outermost layer, and the porous structure of the middle layer shown in Fig. 1. Values such as sound velocity for different layers are listed in Table I. We use k-Wave toolbox [14] in MATLAB for the PA simulation. The pixel size is set to 0.2 mm and the total phantom size is 94.4 mm × 94.4 mm.

TABLE I. THE PARAMETERS SETTING OF SKULL IN PA SIMULATION

|  | Density (kg/m^3) | Wave Speed (m/s) | Attenuation factor (dB/cm/MHz) | Acoustic impedance (MPa • s/m^3) |
|---|---|---|---|---|
| Outer layer | 1900 | 2900 | 8.72 | 5.51 |
| Marrow | 1700 | 2500 | 10.11 | 4.25 |
| Inner layer | 1900 | 2900 | 8.72 | 5.51 |

### B. Simple case simulation results

First, we run a simple PA reflection simulation. The simulation setting is shown in Fig. 2. A smooth reflector is put at the bottom, and a point source in red is assigned. A linear array ultrasound probe in white is put towards the reflecting direction. A typical received PA signal is shown in Fig. 3. Two signal peaks can be found, the first one is received directly, and the second one is received through reflection. Since the sensor is directional to the reflector, the reflective PA signal has a more complete shape.

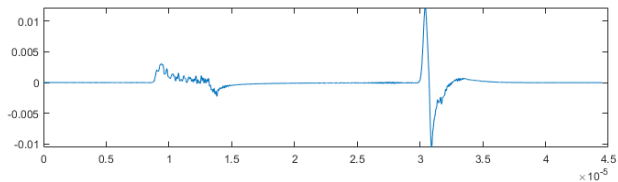

Fig. 3. The signal received of the linear sensor. The first signal is received directly from the source, and the second one is received through reflection.

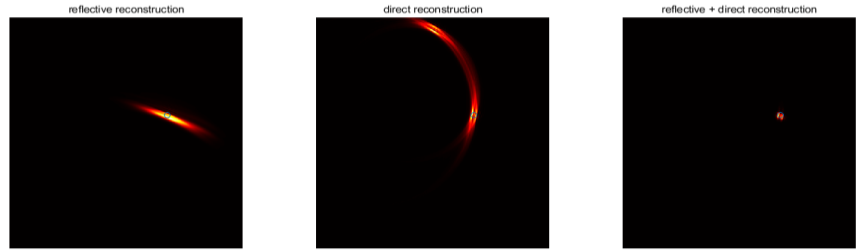

Fig. 4. The reconstruction result of a point source in the simple setting, where an ideal reflector is set. The images from left to right are reflection reconstruction, direct reconstruction, and superposition of both respectively.

The image reconstruction result is shown in Fig. 4. A hollow blue circle is used to represent the source point (ground truth). It can be found that with only reflective PA signals, the point source is reconstructed to be like an arc that is perpendicular to the direction of reflection. On the other hand, with only direct signal, the reconstruction result is also like an arc centered on the linear array. With both PA signals used for image reconstruction, the point source is recovered much better than any one of them.

*C. Human skull simulation results*

We then perform the simulation on human skull phantom. Three linear-array acoustic sensors are put on different directions, which is shown in Fig. 1. The top left one is denoted as sensor 1 that is used for reflective PA signal detection. The top right one is sensor 2 that is used to reconstruct the image by PA signals directly going through the skull. The bottom left sensor is sensor 3, which is on the opposite side of sensor 1, is used for comparison study. Two sets of results are shown in Fig. 5 and Fig. 6.

In Fig. 5, a point source is assigned. We could see that although sensor 2 is directly facing the source, the reconstruction result is strongly affected by the skull leading to inaccurate positioning. Compared with non-line-of-sight PA image with reflective PA signals, the combined reflective and direct reconstruction can improve the image quality greatly. Compared with the result using sensor 3, the result of our proposed method is more similar with the ground truth, instead of an arc. In Fig. 6, a line source is set. It would be difficult for a linear array to reconstruct it due to the limited perspective, but the superimposed result is able to reconstruct the source.

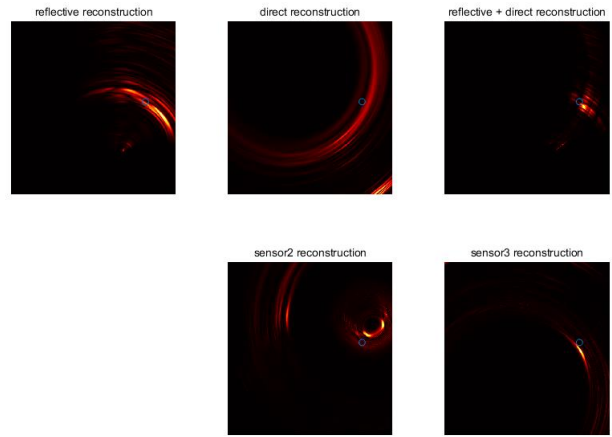

Fig. 5. The reconstruction result comparison of a point source. The first row of images are from reflection reconstruction, direct reconstruction, and superposition of both, respectively. The second row of images come from other two perspectives for comparison.

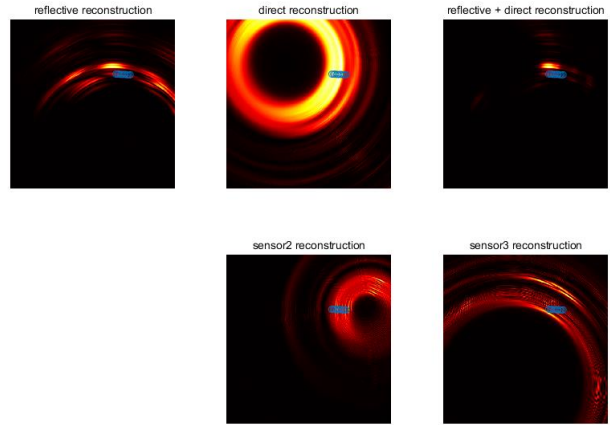

Fig. 6. The reconstruction result comparison of a line source. The first row of images are from reflection reconstruction, direct reconstruction, and superposition of both, respectively. The second row of images come from other two perspectives for comparison.

## IV. Discussion and conclusion

In this work, we propose non-line-of-sight PA imaging method, which uses the strong reflection of ultrasonic waves on the inner wall of the skull to receive the PA signals outside the line-of-sight. The temporal bone window is chosen for PA signal detection to minimize the impact of the skull on the signal. After detecting the non-line-of-sight PA signals reflected by the skull, the non-line-of-sight PA image can be recovered through the corresponding signal and image processing algorithm.

Compared with conventional PA imaging techniques, this method uses reflected PA signals to expand the imaging FOV and further improve the imaging quality from the limited imaging perspective in the temporal bone region. It still faces challenges, such as the selection of reflection interface and the calculation of normal vectors, which would affect the reconstruction results. In the future work, ex vivo and in vivo experimental validation will be performed to further verify its feasibility.

## Acknowledgment

This research was funded by National Natural Science Foundation of China (61805139), United Imaging Intelligence (2019X0203-501-02), Shanghai Clinical Research and Trial Center (2022A0305-418-02), and Double First-Class Initiative Fund of ShanghaiTech University (2022X0203-904-04).